\documentclass[runningheads]{llncs}
\usepackage[T1]{fontenc}
\usepackage{graphicx}
\usepackage{amsmath,amssymb,amsfonts}
\usepackage{amstext}

\usepackage{url}

\begin{document}
\title{``It Looks All the Same to Me'': Cross-index Training for Long-term Financial Series Prediction}
%
%
\author{Stanislav Selitskiy \orcidID{0000-0003-1758-0171}}
\authorrunning{S. Selitskiy}
%
\institute{University of Bedfordshire, Park Square, Luton, LU1 3JU, UK\\
\email{stanislav.selitskiy@study.beds.ac.uk}
}
\maketitle              
\begin{abstract}
We investigate a number of Artificial Neural Network architectures (well-known and more ``exotic'') in application to the long-term financial time-series forecasts of indexes on different global markets. The particular area of interest of this research is to examine the correlation of these indexes' behaviour in terms of Machine Learning algorithms cross-training. Would training an algorithm on an index from one global market produce similar or even better accuracy when such a model is applied for predicting another index from a different market? The demonstrated predominately positive answer to this question is another argument in favour of the long-debated Efficient Market Hypothesis of Eugene Fama.
\end{abstract}
\section{Introduction}
The Efficient Market Hypothesis (EMH) in the well-known form was popularized by Eugene Fama \cite{fama1970efficient} in the late 60's - the early '70s, though the earliest documented high-level formulation was known since 16'th century \cite{sewell2011history}. In various forms of strictness, it postulates that all potentially available information is immediately incorporated into market prices. This quite radical and superficial (and, in a way, superstitious and superfluous) proposition was fiercely debated over time. The early discussion of the EMH hypothesis was primarily conducted by proponents of the hypothesis on the material of the developed markets, which led to over-confident conclusions. In the '80s, the application of the theory was tested on emergent markets, which brought mixed results. In the '90s, the challenge of the EMH became widespread, though, even in the 21st century, the discussion still continues \cite{Yen2008}.

The economic \cite{roll1994every}, environment \cite{roll1984orange}, logical paradoxes \cite{jensen1968performance}, and psychological \cite{shleifer2000inefficient} aspects of the debate are out of the scope of this paper. Machine Learning (ML) and statistical market prediction methods based on the previous historical data are the ``bread and butter'' of the practising traders and one of the arguments favouring the EMH \cite{venugopal1994neural}. However, even though these algorithms have some limited predicted power in plain form, they do not strictly show the use of the information in the predictions but possibly stochastic correlations. Research on tracing how publicly unknown information finds its way to the public and into market prices, the rate and cost of its dissipation is still to be conducted.

In this research, we still use the indirect method. We investigate the hypothesis that if today's global markets are affected by information, then their behaviour is correlated long-term. It is a trivial proposition, which has been debated for decades \cite{eun1989international}, but we give it a practical spin in our applied research, going beyond usual statistics correlation analysis \cite{poterba1988mean}. If we train an ML model on one index (potentially on one market) and then use it for index prediction of another index on another global market, and those predictions have similar or better accuracy (due to less over-fitting), then this correlation would be an argument in favour to EMH. 

In the study, we experiment with various Artificial Neural Network (ANN) architectures, the well-known easily assembled from the layers available in many ML libraries, more ``exotic'' that require additional manual coding, and developed from scratch by the authors, to ensure that cross-training effects are common and not bound to one particular ANN architecture. We show how successfully those ANN models work in long-term forecasting for $30$ days when trained on the same index data and then when they are cross-trained. We use historical data from $2005$ to the beginning of $2022$ for the NASDAQ, Dow Jones, NIKKEI and DAX indexes to cover intra-market and cross-market effects.

The contribution is organized in the following way: Section~\ref{sec:rel_work} presents closely related to the study's existing work, Section~\ref{sec:met} high-level describes ML algorithms chosen for the study, data sets and their partition, which were used in computational experiments, as well as accuracy metrics for algorithms' evaluation. Section~\ref{sec:ex} lists hardware parameters and model configurable parameters, as well as details of the less-popular ANN architectures implemented from scratch. Section~\ref{sec:res} shows the results of the experiments in diagram and table form, discusses these results and study limitations, draws conclusions and outlines future research directions.

\section{Related work}
\label{sec:rel_work}
ML algorithms of different types were used for various markets, stock types, observation periods, and prediction intervals.

Two classes of the ML algorithms, Decision Tree (DT) based and ANN-based, were used on the $10$ years interval for the Tehran Stock Exchange data (financial, petroleum, mineral, and metal indexes) in \cite{nabipour2020deep}. Various prediction intervals ranging from $1$ to $30$ days in the future were used, measuring accuracy with four metrics. Unfortunately, the best results were reported over the whole time span and prediction intervals, which makes the research of limited interest. However, systematically, Long Short-term Memory (LSTM) ANN was reported as superior to other ML models.

In \cite{stoean2019deep}, not just composite indexes, but instead, stocks of the $25$ individual companies on the Bucharest Stock Exchange were experimented with on the span of over $21$ years. Two ANN models, LSTM and $1$-dimensional temporal Convolutional Neural Network (CNN), were used not just for accuracy calculation but also for the trading gain simulation in comparison with traditional simple trading tactics. Predictions were run on training intervals from $30$ to $120$ days, but with only $1$ day prediction in the future at a time. Both architectures demonstrated superior performance compared to the simple tactics specialising in the window frequency for CNN or gain amounts for LSTM. 

Traditionally ML models are trained on the same indexes but on the previous chronological data. Of our particular interests are approaches of the ML models training on the different
time series. ``Cross-training'' is not a well-established term; therefore, such methods come under various rubrics, such as ensembles, pre-training, and transfer learning. They include noise-augmented training data \cite{ZHANG2007}, contrastive self-supervised learning targeting on the extraction of the augmented components \cite{WICKSTROM2022}, averaged LSTM training on multiple index data \cite{tsang2018recurrent}, or averaging of the ensemble of the models trained on the multiple indexes \cite{he2020multi}.

Although some similarities could be seen between the above-mentioned and presented research, it is worth pointing out that the aim of the existing research is the enriching training data of the pre-trained transfer-learning models with the superset of the patterns found either in other indexes (or domain) or synthetic time series. In our study, ANN models are trained on the index of one single market and then tested on the other three markets, and such experiments are applied to all four markets in a circular fashion. No averaging or other use of data from multiple markets for training is done because the study aims to experimentally confirm the alleged by EMH synchronicity of the markets, strong enough to have no need for averaging or multiple data sets pre-training.

\section{Methods}
\label{sec:met}

\begin{figure}
  \centering

\begin{minipage}{1.0\linewidth}
  \centering
  \centerline{
  \includegraphics[width=0.24\linewidth]{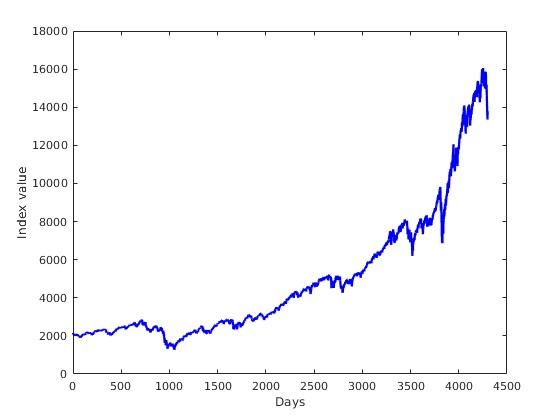}
  \includegraphics[width=0.24\linewidth]{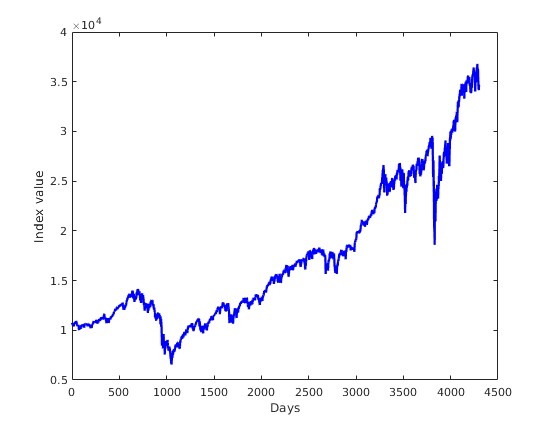}

  \includegraphics[width=0.24\linewidth]{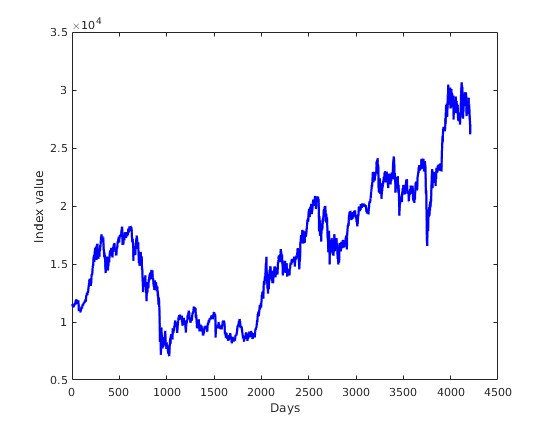}
  \includegraphics[width=0.24\linewidth]{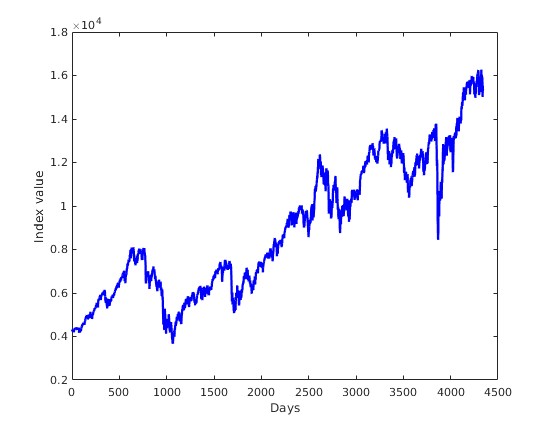}
  }
\end{minipage}

\caption{Stock indexes NASDAQ, DOW, NIKKEI, DAX on intervals of the experiments.}
\label{fig:index}
\end{figure}

It could be easily seen in Figure~\ref{fig:index}, 
Table~\ref{p.corr}, the stock indexes used in this study and described in more detail in Section~\ref{sec:ds}, are correlated. 

%
%

\begin{table}
\caption{Pearson correlation coefficient, applied pairwise to the stock indexes under experimentation}
\label{p.corr}
\begin{tabular}{|l|l|l|}
\hline
Index 1 & Index 2& Correlation\\
\hline
NASDAQ & NIKKEI & 0.8879\\
NASDAQ & DOW & 0.9750\\
NASDAQ & DAX & 0.8975\\
DOW & DAX & 0.9456\\
DOW & NIKKEI & 0.9053\\
NIKKEI & DAX & 0.8740\\
\hline
\end{tabular}
\end{table}

However, how much of this correlation can be used for practical purposes of cross-training and cross-forecasting? Furthermore, how much of theoretical conclusions in the context of EMH can we extract? Remaining inside statistical methods, such cross-training is not intuitive, especially from the Frequentist's point of view, though the Bayesian can look at statistics obtained from another index as a starting belief to polish them on the target stock statistical forecasts using statistical \cite{jakaite2012bayesian,schetinin2011informativeness,SelitskiyLOD2022} or ML methods \cite{SelitskiyAGI2023}. Also, choosing particular statistics requires prior assumptions about patterns that can embed information the stocks are supposed to absorb by EMH.

ML methods, and especially ANNs', strength is freedom of necessity for such \textit{a priory} assumptions. ANN can learn unexpected patterns, and if there are such patterns in the stock series which can be cross-learned and successfully used for cross-forecasting, that is an argument for EMH, considering the ubiquity of the information in global markets, economy, and the world itself. To leverage the best models mentioned in Section~\ref{sec:rel_work} and enhance them, we concentrated on ANN architectures, significantly increasing the variety of the being studied models and applying them for the extreme for the cited works prediction interval of $30$ days.

Accuracy metrics (described below in Subsection~\ref{sec:am}) and their distribution over session partitions (also described below in Subsection~\ref{sec:ds}), are collected for all being investigated ANN models. As ablation testing, prediction accuracies are calculated in the absence of cross-training (the same index is used for training and testing, but on different time intervals). Acquired accuracy distributions for cross-training are subjected to the non-parametric hypothesis testing Wilcoxon signed rank algorithm to verify that non-cross-training and cross-training accuracy distribution either have no shift or ``right'' (greater) shift for non-cross-training distributions, i.e. cross-training accuracy is at least no worse than the non-cross-training.

\subsection{Accuracy metrics}
\label{sec:am}
As an accuracy metrics, we use the Mean Absolute Percentage Error (MAPE), defined as follows:
\begin{equation}
MAPE=\frac{1}{n}\sum_{t=1}^n |\frac{A_t-F_t}{A_t}|
\end{equation}
And Root Mean Square Error (RMSE): 
\begin{equation}
RMSE=(\frac{1}{n}\sum_{t=1}^n{(A_t-F_t)^2})^{\frac{1}{2}}
\end{equation}

Where $A_t$ and $F_t$ are the actual and predicted indexes at a given day $t$, respectively, and $n$ is the number of test observations.

\subsection{Data sets}
\label{sec:ds}
The data for NASDAQ, Dow Jones, NIKKEI, and DAX indexes for the beginning of $2005$ year up to the end of January $2022$ were used in computational experiments. The data were downloaded from \url{https://tradingeconomics.com}. Detailed statistical analysis of the data set is out of the scope of the paper; however, general intuition about this strongly non-stationary time series can be obtained from general Figure~\ref{fig:index}, 
and simple Pearson correlation coefficients Table~\ref{p.corr}. The data represent a broad spectrum of behavioural tendencies: oscillations of the stagnant markets, explosive growth and fall during bubble bursts and busts in American, European, and Asian markets. The difference in stock exchanges' working days schedules made the data slightly (a few days) asynchronous. The reason for the particular indexes selection is an attempt to limit the combinatorial complexity of the experiments (only four indexes), still preserving geographical representability (at least for the most economically important markets: DOW, NIKKEI, DAX), and general vs. industry-specific markets (DOW and NASDAQ).

\begin{figure}
  \centering
  \includegraphics[width=1.0\linewidth]{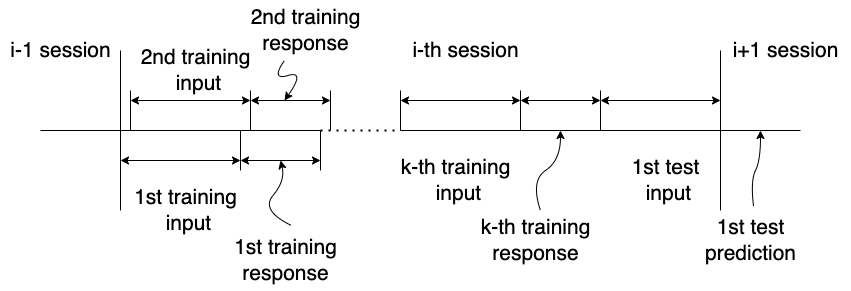}
\caption{Session partition schema.}
\label{fig:sess.part}
\end{figure}

To accommodate long-term prediction of $30$ days forward, based on the past $30$ days indexes performance, the data were divided into $35$ subsets, each consisting of $30$ observations used to train the sequence-to-sequence models. Each ``observation" was comprised of $30$ days values and the output ``label'' values were the $31^{th}-60^{th}$ days. Each following observation starts with a $1$ day shift. The number of training sessions is first $34$, and the whole session is $120$ days. It was defined so that none of the training data (including the label days) would touch the following test session data. The number of test sessions is last $34$. The last $30$ days of the preceding training session (not used in the training process) were used to predict the first $30$ days of the following test session, with step $1$ until the end of the test session. Models' parameters were reset for each session, and training was done anew (see \ref{fig:sess.part}). For sequence-to-value models, such as LSTM, the whole $119$ + $1$ last label day was used in the training session.

\section{Experiments}
\label{sec:ex}

The experiments were run on the Linux (Ubuntu 20.04.3 LTS) operating system with two dual Tesla K80 GPUs (with $2\times 12$GB GDDR5 memory each) and one QuadroPro K6000 (with $12$GB GDDR5 memory, as well), X299 chipset motherboard, 256 GB DDR4 RAM, and i9-10900X CPU. Experiments were run using MATLAB 2022a with Deep Learning Toolbox.
For inferential statistics, built-in R 4.2.1 implementations were used with default parameters unless mentioned otherwise.

\begin{table}
\caption{Accuracy of various ML models for NASDAQ index}

\label{tab:1}
\begin{tabular}{|l|l|l|}
\hline
Model & MAPE & RMSE \\
\hline
AR &$0.0521\pm0.0428$&$336.79\pm471.61$\\
ANN reg. &$0.0504\pm0.0401$&$320.75\pm405.89$\\
Logistic &$0.0541\pm0.0417$&$313.21\pm315.77$\\
ReLU &$0.0560\pm0.0421$&$361.88\pm432.68$\\
LSTM vec. &$0.0506\pm0.0273$&$283.15\pm229.51$\\
LSTM &$0.0615\pm0.0590$&$316.60\pm274.46$\\
SCNN &$0.0643\pm0.0590$&$360.77\pm358.64$\\
RBF &$0.0588\pm0.0418$&$343.20\pm336.99$\\
KGate &$0.0623\pm0.0555$&$363.56\pm351.92$\\
GMDH &$0.0549\pm0.0505$&$316.23\pm321.70$\\
\hline
\end{tabular}
\end{table}

The following models were experimented with: analytically-solved multi-variate linear Auto-regression, ANN Regression (no activation functions), ANN with ReLU (Rectified Linear Unit), Logistic, Hyperbolic tangent activations, sequence-to-sequence and sequence-to-value LSTM and GRU (Gated Recurrent Unit), simple sequential and spectral cascade CNN, RBF (Radial Basis Function), KGate (Kolmogorov's Gate), and finally GMDH (Group Method of Data Handling) ANNs. Because ANN regression, ANN with ReLU, KGate activations and CNNs are tolerable to non-normalized input data and frequently produce better accuracy, for these models, experiments are done with non-normalized input. For other models sensitive to normalization, a min-max normalization is applied in a strict mode, calculated only for a given observation data without looking ahead of time or in the past. A mean squared error was used as a loss function for all ANNs.

\subsection{Ready-to-use ANN layers}

Well-known ANN layers and architectures, readily available in MATLAB or its Toolboxes, are not described here in detail - only non-default configuration parameters are presented below. Non-standard implementations and the whole source code used in experiments can be found on GitHub (\url{https://github.com/Selitskiy/LTTS}) and are implemented as follows.

If we look at linear regression, or the linear part of a layer transformation of an ANN, as a transformation from a higher dimensional space into the lower dimensional space $f, \, n<m$:

\begin{equation}
f:\mathcal{X}\subset \mathbb{R}^m \mapsto \mathcal{Y} \subset \mathbb{R}^n
\label{eq:3}
\end{equation}

Linear regression can be represented as a matrix multiplication:

\begin{equation}
\textbf{y} = f(\textbf{x}) = \textsf{W}\textbf{x}, \, \forall \textbf{x} \in \mathcal{X}\subset \mathbb{R}^m, \, \forall \textbf{y} \in \mathcal{Y} \subset \mathbb{R}^n
\label{eq:4}
\end{equation}
where $\textsf{W} \in \mathcal{W} \subset \mathbb{R}^{n \times m}$ is the adjustable coefficient matrix that, using minimization of the sum of squared errors, can be found as \cite{hastie01statisticallearning}:

\begin{equation}
\textbf{W} = (\hat{\textbf{X}}^T \hat{\textbf{X})}^{-1} \hat{\textbf{X}}^T \hat{\textbf{Y}}
\label{eq:5}
\end{equation}
where $\hat{\textbf{X}}, \, \hat{\textbf{Y}}$ are matrices of the observations of the input and output observations, respectively \cite{selitskaya2020deep}.

All other ANN architectures, to compare them on a similar level of complexity (number of learnable parameters), were designed to have two hidden layers with a number of neurons $m$ and $2m+1$ on the first and second hidden layer, respectively, where $m$ is an input dimensionality, Formula~\ref{eq:3}. The reason to limit ANN models to two layers was if one looks at ANN as a Universal Approximation according to the Kolmogorov-Arnold superposition theorem \cite{kolmogorov1961representation}, for general emulation of the $f:\mathcal{X}\subset \mathbb{R}^m \mapsto \mathcal{Y} \subset \mathbb{R}$ process, the $2$-layer is a minimal ANN configuration needed (given that activation functions are complex enough):

\begin{equation}
\label{eq:6}
f(\textbf{x}) = f(x_1, \dots , x_m) = \sum_{q=0}^{2m} \Phi_q (\sum_{p=1}^{m} \phi_{qp}(x_p))
\end{equation}
where $\Phi_q$ and $\phi_{qp}$ are continuous $\mathbb{R} \mapsto \mathbb{R}$ functions.

The practicality of such ANN, as a Universal Approximator, was disputed in \cite{girosi1989representation}, particularly because of the non-smoothness, hence non-practicality, of the inner $\phi_{qp}$ functions. However, these objections were rebutted in \cite{Kurkova1991}. In \cite{pinkus1999approximation} $\phi_{qp}$ activation functions are even called ``pathological''.

Therefore, in addition to the usual activation functions, much less frequently used architectures and activation functions were experimented with.

\begin{table}
\caption{Accuracy of various ML models for NIKKEI index}
\label{tab:3}
\begin{tabular}{|l|l|l|}
\hline
Model & MAPE & RMSE \\
\hline
AR & $0.0808\pm0.0723$ & $1919.10\pm2707.51$ \\
ANN reg. &$0.0799\pm0.0723$ &$1877.54\pm2630.55$\\
Logistic &$0.0890\pm0.0798$ &$1963.34\pm2596.88$\\
ReLU &$0.0904\pm0.0989$&$2256.28\pm3905.08$\\
LSTM vec. &$0.0598\pm0.0248$&$1165.07\pm467.01$\\
LSTM &$0.0587\pm0.0286$&$1120.97\pm519.30$\\
SCNN &$0.0932\pm0.0924$&$2122.16\pm3200.82$\\
RBF &$0.0923\pm0.0818$&$2105.33\pm3015.20$\\
KGate &$0.1008\pm0.1086$&$2386.27\pm3854.98$\\
GMDH &$0.0932\pm0.0926$&$2087.34\pm2806.60$\\
\hline
\end{tabular}
\end{table}

\subsection{Custom-coded and originally-developed ANN layers}

Radial Basis Functions (RBF) ANN were proposed at the end of the 80s \cite{broomhead1988radial}. It could be viewed as a ``soft gate'' which activates the transformation matrix coefficients in Gaussian proportion to the proximity of the test signal to the training signals, this transformation matrix's coefficients were trained at \cite{Park1991}. 

\begin{equation}
\label{eq:7}
f(\textbf{x}) = \textbf{a}_k e^{-\textbf{b}_k(\textbf{x}-\textbf{c}_k)^2}
\end{equation}
where $k \in \{1 \dots n\}$.

An apparent drawback of the architecture is its ``fluffiness'' due to the non-reuse of the neurons for the ``missed'' test-time data input, making the RBF ANNs less dense or compact than Deep ReLU ANNs. Still, RBF is a viable architecture used in niche applications \cite{kurkin2018artificial,beheim2004new}.

Another ANN architecture \cite{Selitskiy2022} can be seen as a part of the Gated Linear Unit (GLU) family of activations. Using Directed Acyclic Graph (DAG) ANN, one can implement a cell (let us call it Kolmogorov's Gate or KGate for short) of perceptrons with logistic sigmoid activations that would work as allow or do not allow gates at saturation domain or multiplicative scaling of the main trunk of ANN, in the non-saturation domain of input values. Perceptrons with hyperbolic tangent activation would work as update/forget or the mean shift gates on the main ANN trunk, working together with the linear input transformation through the multiplication gate, Formula~\ref{eq:8}.

\begin{equation}
\label{eq:8}
\begin{split}
\textbf{z}_i &= (\textsf{W}_i\textbf{x}_i + (\tau \circ \textsf{W}_{ti}\textbf{x}_0) \odot (\textsf{W}_{ai}\textbf{x}_0)) \odot \sigma \circ \textsf{W}_{si}\textbf{x}_0, \, \\ \forall &\textbf{x}_0 \in \mathcal{X}_0 \subset \mathbb{R}^m, \, \forall \textbf{x}_i \in \mathcal{X}_i \subset \mathbb{R}^{m_i}
\end{split}
\end{equation}

where $\textbf{x}{_0}$ is an ANN input, $\textbf{x}_i$ is an input of the $i^{th}$ layer, $\textsf{W}_i\textbf{x}_i$ is the linear transformation of the main trunk, $\textsf{W}_{ti}\textbf{x}_0, \, \textsf{W}_{ai}\textbf{x}_0, \, \textsf{W}_{si}\textbf{x}_0$ are linear transformations inside the KGate cell, and $\tau, \, \sigma$ are hyperbolic tangent and logistic sigmoid activation functions, respectively.

Following Ivakhnenko \cite{ivakhnenko1971polynomial}, the multi-layer neural-network models could be grown by the Group Method of Data Handling (GMDH) using a neuron activation function defined by a short-term polynomial, in our case - the polynomial of the second degree of the pairwise connected neurons, which ensures linear gradient optimization hyperplane on each layer generation step \cite{nyah2016evolving}. 

\begin{equation}
\label{eq:9}
f(x_i, x_j) = (w_{ki}x_i + w_{kj}x_j + w_{k0})^2
\end{equation}

The GMDH is capable of generating new layers capable of predicting new data most accurately. The GMDH generates new neurons to be fitted to the training data in each layer. A given number of the best-fitted neurons are selected for the next layer.

Spectral cascade CNN (SCNN) was organized as parallel Directed Acyclic Graph (DAG) cells, similar to Inception or ResNet type cells \cite{szegedy2015going,ren2016deep}, of $3 \times 1$, $5 \times 1$, $7 \times 1$, $11 \times 1$, $13 \times 1$ dimensions, packets of $16$ of each \cite{SelitskiyMVSK2022}.

ANN models were trained using the ``adam'' learning algorithm with $0.01$ initial learning coefficient, mini-batch size $32$, and $1000$ epochs.

\section{Results}
\label{sec:res}

As mentioned above, computational experiments were conducted on NASDAQ, Dow Jones, NIKKEI, and DAX indexes partitioned as described in Section~\ref{sec:ds} using linear Auto-regression, ANN Regression, ANN with ReLU, Logistic, Hyperbolic tangent activations, sequence-to-sequence and sequence-to-value LSTM and GRU, simple sequential and spectral cascade CNN, RBF, KGate, Transformer-like non-linearity, and GMDH ANNs.

\begin{table}
\caption{MAPE of cross-trained LSTM sequence-to-vector model (rows - trained on the index, columns - tested on the index).}
\label{tab:lstmsv.a}
\begin{tabular}{|l|l|l|l|l|}
\hline
Index & NASDAQ & DOW & NIKKEI & DAX\\
\hline
NASDAQ & $0.0506\pm0.0273$ & $0.0409\pm0.0269$ & $0.0574\pm0.0258$ & $0.0523\pm0.0239$ \\
DOW & $0.0501\pm0.0261$ & $0.0417\pm0.0270$ & $0.0589\pm0.0280$ & $0.0546\pm0.0268$ \\
NIKKEI & $0.0521\pm0.0284$ & $0.0416\pm0.0243$ & $0.0598\pm0.0248$  & $0.0551\pm0.0243$ \\
DAX & $0.0522\pm0.0291$ & $0.0425\pm0.0283$ & $0.0615\pm0.0291$ & $0.0589\pm0.0377$ \\
\hline
\end{tabular}
\end{table}

Logistic and Hyperbolic tangent activation ANNs, LSTM and GRU, KGate and Transformer, CNN and SCNN produced similar results; therefore, only one of them is shown. For ablation non-cross-training accuracy data, only NASDAQ and NIKKEI index results are shown as extremes. To save space, Dow Jones and DAX results are not shown here 
Table~\ref{tab:1}, Table~\ref{tab:3}), but can be downloaded from the same GitHub link as the source code.

\begin{table}
\caption{MAPE of cross-trained RBF model (rows - trained on the index, columns - tested on the index).}
\label{tab:rbf.a}
\begin{tabular}{|l|l|l|l|l|}
\hline
Index & NASDAQ & DOW & NIKKEI & DAX \\
\hline
NASDAQ & $0.0588\pm0.0418$ & $0.0485\pm0.0393$ & $0.0756\pm0.0397$ & $0.0590\pm0.0316$ \\
DOW & $0.1328\pm0.4250$ & $0.1020\pm0.3150$ & $0.0709\pm0.0344$ & $0.4299\pm2.1876$ \\
NIKKEI & $0.0698\pm0.0680$ & $0.0544\pm0.0619$ & $0.0923\pm0.0818$ & $0.0592\pm0.0354$ \\
DAX & $0.0649\pm0.0503$ & $0.0542\pm0.0521$ & $0.0702\pm0.0339$ & $0.0794\pm0.1053$ \\
\hline
\end{tabular}
\end{table}

Results of the cross-training experiments are shown for all indexes, but also for extremes of the models - LSTM sequence-to-vector being most accurate, and RBF activation being most vulnerable for volatility: Table~\ref{tab:lstmsv.a}, Table~\ref{tab:rbf.a}.

P-values of the paired Wilcoxon signed rank test between non-cross-trained and cross-trained accuracy distributions for various ANN architectures for NASDAQ, Dow Jones, NIKKEI indexes, Table~\ref{tab:p-nasdaq}, \ref{tab:p-dow}, \ref{tab:p-nikkei}. Again, to save space, DAX results, which are similar to NASDAQ, are not shown.

\begin{table}
\caption{Wilcoxon signed rank test p-values on accuracy distribution over $34$ sessions for models trained on NASDAQ data and tested on other indexes. P-values for alternative hypotheses of distribution shift for the not-cross-trained relative to cross-trained to the right, left, and two-sided.}
\label{tab:p-nasdaq}
\begin{tabular}{|l|l|l|l|l|}
\hline
Model & Index & Greater & Less & Two-side\\
\hline
Reg & DOW & 0.00003 & 0.99997 & 0.00007 \\
Reg & NIKKEI & 0.97359 & 0.02757 & 0.05513 \\
Reg & DAX & 0.89874 & 0.10434 & 0.20869 \\
ANN & DOW & 0.00003 & 0.99997 & 0.00006 \\
ANN & NIKKEI & 0.97683 & 0.02421 & 0.04843\\
ANN & DAX & 0.90761 & 0.09528 & 0.19056 \\
ReLU & DOW & 0.00001 & 0.99999 & 0.00001 \\
ReLU & NIKKEI & 0.93946 & 0.06276 & 0.12552\\
ReLU & DAX & 0.53362 & 0.47310 & 0.94619\\
Sig & DOW & 0.00160 & 0.99850 & 0.00319\\
Sig & NIKKEI & 0.85610 & 0.14799 & 0.29599\\
Sig & DAX & 0.81975 & 0.18476 & 0.36953\\
LSTMv & DOW & 0.00000 & 0.99999 & 0.00000\\
LSTMv & NIKKEI & 0.86407 & 0.13988 & 0.27976\\
LSTMv & DAX & 0.68786 & 0.31815 & 0.63630\\
LSTM & DOW & 0.00004 & 0.99996 & 0.00009\\
LSTM & NIKKEI & 0.99013 & 0.01038 & 0.02077\\
LSTM & DAX & 0.49327 & 0.51346 & 0.98654\\
GMDH & DOW & 0.00004 & 0.99996 & 0.00009\\
GMDH & NIKKEI & 0.79701 & 0.20805 & 0.41609\\
GMDH & DAX & 0.75591 & 0.24945 & 0.49890\\
KGate & DOW & 0.00003 & 0.99997 & 0.00006\\
KGate & NIKKEI & 0.94161 & 0.06054 & 0.12109\\
KGate & DAX & 0.49327 & 0.51346 & 0.98654\\
RBF & DOW & 0.00069 & 0.99936 & 0.00137\\
RBF & NIKKEI & 0.97683 & 0.02421 & 0.04843\\
RBF & DAX & 0.55369 & 0.45299 & 0.90598\\
SCNN & DOW & 0.00000 & 0.99999 & 0.00001\\
SCNN & NIKKEI & 0.76010 & 0.24545 & 0.49089\\
SCNN & DAX & 0.61937 & 0.38708 & 0.77417\\
\hline
\end{tabular}
\end{table}

\section{Discussion and Conclusions}
\label{sec:disc}

Interesting observations for ablation experiments are based on the MAPE metric, giving a more universal measure of the overall accuracy for each prediction point across the multiple indexes. In contrast, RMSE, though more index-specific, penalizes outlier predictions more, even though few of them exist. ANN models with exponential and polynomial non-linearities are especially vulnerable to stock volatility especially observed in NIKKEI and DAX indexes behaviour, which is one of the faces of the Out-of-Distribution (OOD) problem. Among all ANN architectures, the most accurate and robust was LSTM sequence-to-vector architecture.

\begin{table}
\caption{Wilcoxon signed rank test p-values on accuracy distribution over $34$ sessions for models trained on DOW data and tested on other indexes. P-values for alternative hypotheses of distribution shift for the not-cross-trained relative to cross-trained to the right, left, and two-sided.}
\label{tab:p-dow}

\begin{tabular}{|l|l|l|l|l|}
\hline
Model & Index & Greater & Less & Two-side\\
\hline
Reg & NASDAQ & 0.99988 & 0.00013 & 0.00027\\
Reg & NIKKEI & 0.99993 & 0.00007 & 0.00015\\
Reg & DAX & 0.99948 & 0.00056 & 0.00112\\
ANN & NASDAQ & 0.99979 & 0.00023 & 0.00047\\
ANN & NIKKEI & 0.99996 & 0.00005 & 0.00009\\
ANN & DAX & 0.99952 & 0.00052 & 0.00104\\
ReLU & NASDAQ & 0.99980 & 0.00022 & 0.00043\\
ReLU & NIKKEI & 0.99995 & 0.00006 & 0.00012\\
ReLU & DAX & 0.99977 & 0.00025 & 0.00050\\
Sig & NASDAQ & 0.99940 & 0.00064 & 0.00128\\
Sig & NIKKEI & 0.99854 & 0.00156 & 0.00311\\
Sig & DAX & 0.99272 & 0.00766 & 0.01531\\
LSTMv & NASDAQ & 0.99999 & 0.00001 & 0.00001\\
LSTMv & NIKKEI & 0.99993 & 0.00008 & 0.00016\\
LSTMv & DAX & 0.99999 & 0.00001 & 0.00002\\
LSTM & NASDAQ & 0.99999 & 0.00001 & 0.00002\\
LSTM & NIKKEI & 0.99952 & 0.00051 & 0.00103\\
LSTM & DAX & 0.98755 & 0.01304 & 0.02609\\
GMDH & NASDAQ & 0.99830 & 0.00181 & 0.00361\\
GMDH & NIKKEI & 0.99956 & 0.00048 & 0.00095\\
GMDH & DAX & 0.94564 & 0.05631 & 0.11262\\
KGate & NASDAQ & 0.99992 & 0.00009 & 0.00018\\
KGate & NIKKEI & 0.99995 & 0.00006 & 0.00012\\
KGate & DAX & 0.99940 & 0.00064 & 0.00128\\
RBF & NASDAQ & 0.99999 & 0.00001 & 0.00003\\
RBF & NIKKEI & 0.99742 & 0.00274 & 0.00548\\
RBF & DAX & 0.99272 & 0.00766 & 0.01531\\
SCNN & NASDAQ & 0.99993 & 0.00008 & 0.00015\\
SCNN & NIKKEI & 0.99969 & 0.00033 & 0.00065\\
SCNN & DAX & 0.99019 & 0.01029 & 0.02058\\
\hline
\end{tabular}
\end{table}

\begin{table}
\caption{Wilcoxon signed rank test p-values on accuracy distribution over $34$ sessions for models trained on NIKKEI data and tested on other indexes. P-values for alternative hypotheses of distribution shift for the not-cross-trained relative to cross-trained to the right, left, and two-sided.}
\label{tab:p-nikkei}
\begin{tabular}{|l|l|l|l|l|}
\hline
Model & Index & Greater & Less & Two-side\\
\hline
Reg & NASDAQ & 0.00137 & 0.99872 & 0.00273\\
Reg & DOW & 0.00009 & 0.99991 & 0.00019\\
Reg & DAX & 0.01536 & 0.98536 & 0.03071\\
ANN & NASDAQ & 0.00128 & 0.99881 & 0.00255\\
ANN & DOW & 0.00007 & 0.99993 & 0.00015\\
ANN & DAX & 0.01851 & 0.98233 & 0.03701\\
ReLU & NASDAQ & 0.00128 & 0.99881 & 0.00255\\
ReLU & DOW & 0.00009 & 0.99992 & 0.00017\\
ReLU & DAX & 0.00685 & 0.99351 & 0.01369\\
Sig & NASDAQ & 0.00017 & 0.99984 & 0.00034\\
Sig & DOW & 0.00001 & 0.99999 & 0.00001\\
Sig & DAX & 0.00097 & 0.99909 & 0.00195\\
LSTMv & NASDAQ & 0.04666 & 0.95510 & 0.09332\\
LSTMv & DOW & 0.00009 & 0.99991 & 0.00019\\
LSTMv & DAX & 0.15642 & 0.84783 & 0.31283\\
LSTM & NASDAQ & 0.36879 & 0.63785 & 0.73757\\
LSTM & DOW & 0.01148 & 0.98908 & 0.02295\\
LSTM & DAX & 0.25674 & 0.74894 & 0.51348\\
GMDH & NASDAQ & 0.00214 & 0.99799 & 0.00428\\
GMDH & DOW & 0.00038 & 0.99965 & 0.00076\\
GMDH & DAX & 0.01396 & 0.98670 & 0.02791\\
KGate & NASDAQ & 0.00112 & 0.99896 & 0.00223\\
KGate & DOW & 0.00001 & 0.99999 & 0.00002\\
KGate & DAX & 0.00026 & 0.99976 & 0.00051\\
RBF & NASDAQ & 0.00177 & 0.99834 & 0.00354\\
RBF & DOW & 0.00003 & 0.99997 & 0.00007\\
RBF & DAX & 0.00035 & 0.99967 & 0.00070\\
SCNN & NASDAQ & 0.01536 & 0.98536 & 0.03071\\
SCNN & DOW & 0.00000 & 1.00000 & 0.00000\\
SCNN & DAX & 0.00097 & 0.99909 & 0.00195\\
\hline
\end{tabular}
\end{table}

It also should be noted that this study is not about transfer learning - the approach popular in image recognition, when an ANN model trained on as much as possible large data is slightly modified and rapidly retrained for recognition of images from other domains, because it has already learned common image micro-patterns. That approach could have been used for stock indexes when time series from other domains, for example, weather prediction, are retrained specifically for financial series. However, that is not the aim of the study - we principally do not allow any retraining, hoping to find synchronous information-shared invariants in the untouched cross-trained models.

Results of the Wilcoxon tests (Tables~\ref{tab:p-nasdaq}-\ref{tab:p-nikkei}) look mixed, with cross-trained accuracy being consistently better across all ANN models for NIKKEI index, consistently worse across all ANN models for Dow Jones, and mixed for NASDAQ and DAX. However, the results are shown for $0$ expected shift. If the shift is bounded by $0.02$ (which is less than the standard deviation of even the best and narrowest accuracy distributions), then all models and stocks demonstrate at least no worse accuracy of the cross-trained models compared to the ablation non-cross-trained ones.

As a conclusion for practical purposes, if the bounded to the circa standard deviation, degradation of the cross-trained modes is accepted, in such a Lipschitz sense, the study supports a weak form of EMH, indirectly supporting the idea of existing of the patterns common for global markets and different stock types, which is allegedly maps to information in EMH context, of course in the study limitations frames. Another observation arguing in favour of the information nature of EMH is the tendency of larger indexes to ``explain'' smaller indexes, i.e. NIKKEI and NASDAQ vs DOW and DAX.

From the general ML point of view, considering local stock indexes as observations of the global process (world market), this study offers experimental validation of the plausible, but still speculations, that ML model trained on some local observations may effectively predict other synchronous observations.

\subsection{Limitations}
The study has obvious limitations in the geographic coverage and representation of only developed economies' stock exchanges. An apparent technical limitation of the study is using ANN models only, skipping other ML and statistical approaches. Another technical one, imposed by the hardware limitations, is the relative simplicity of the ANNs being used, barely fitting in the theoretical minimum of the universal approximation. The more subject-related limitation is that accuracy was calculated uniformly along the whole time span, without concentrating on the times of perturbations when a lot of new information allegedly entered the market. 

\subsection{Future Work}
As an area for future research in ablation experiments area, the common for many ML models, rare but catastrophic prediction failures call for relation-aware methods (for example, graph ANN or physics-aware, or rather ``kinematic-aware'' ANN) that would limit sudden changes in the stock prediction in a limited time period. Attention-aware methods, which would look for similar input period models in the past, may also be for future research. 

Such an approach is also methodologically co-located with the abovementioned limitations - the study of the ``disruption information''-rich time intervals, perhaps in contrast with new information-deprived periods. 
For example, the study of the models trained immediately before and tested immediately after the ``disruption information'' boundaries, such as the emergence of the COVID pandemic or the start of the war in Ukraine. ``Information'' here is used in its common sense meaning rather than in the information theory's. 

Because such information may have prolonged inertia to be digested by markets (for example, intelligence or other expert communities may have that information months in advance), an extended temporal horizon ``cross-training'' may be a prospective area of research.

Extension of the research to emergent markets and behaviour of the individual stocks, especially those which do not follow the expected trend (cross-training of the successful company stocks on the failing companies' data and vice versa), is another area of future research.

%
%
%

\bibliographystyle{splncs04}
\bibliography{refs}

\end{document}